\documentclass[prd,11pt]{revtex4-1}

\usepackage{bm}
\usepackage{amsmath}    % need for subequations
\usepackage{graphicx}   % need for figures
\usepackage{verbatim}   % useful for program listings
\usepackage{color}      % use if color is used in text
\usepackage{subfigure}  % use for side-by-side figures
\usepackage{hyperref}   % use forhttps://de.overleaf.com/project/5ddbf14b7a2b8700016cd45a hypertext links, including those to external documents and URLs
\definecolor{amaranth}{rgb}{0.9, 0.17, 0.31}
\definecolor{purple(munsell)}{rgb}{0.62, 0.0, 0.77}
\definecolor{americanrose}{rgb}{1.0, 0.01, 0.24}
\definecolor{palatinateblue}{rgb}{0.15, 0.23, 0.89}
\definecolor{royalblue(web)}{rgb}{0.25, 0.41, 0.88}
\definecolor{hanpurple}{rgb}{0.32, 0.09, 0.98}
\definecolor{beaublue}{rgb}{0.74, 0.83, 0.9}
\definecolor{carminered}{rgb}{1.0, 0.0, 0.22}
\definecolor{brightpink}{rgb}{1.0, 0.0, 0.5}
\definecolor{vividviolet}{rgb}{0.62, 0.0, 1.0}

\hypersetup{ linktoc=all,
    colorlinks, linkcolor={palatinateblue},
    citecolor={brightpink}, urlcolor={amaranth}}

\newcommand{\be}{\begin{equation}}
\newcommand{\ee}{\end{equation}}
\newcommand{\bs}{\begin{split}} 
\newcommand{\bea}{\begin{eqnarray}}
\newcommand{\eea}{\end{eqnarray}}

\begin{document}

\title{ Curvature of $\kappa $-Poincaré and Doubly Special Relativity} 

\author{Nosratollah Jafari}\email{nosrat.jafari@fai.kz}

\affiliation{Fesenkov Astrophysical Institute, 050020, Almaty, Kazakhstan;
\\
Al-Farabi Kazakh National University, Al-Farabi av. 71, 050040 Almaty, Kazakhstan;
\\
 Center for Theoretical Physics, Khazar University, 41 Mehseti Street, Baku, AZ1096, Azerbaijan}

\begin{abstract}

We study the $\kappa$-Poincaré and the Magueijo-Smolin (MS) DSR in the contex of the relative locality theory. This theory assigns connection, torsion and curvature to momentum space of every modified theory beyond special relativity. We obtain these quantities for the $\kappa$-Poincaré and the MS DSR in all order of the Planck length, at the every point of the momentum space. The connection for the $\kappa$-Poincar\'e theory and the MS DSR can be non-zero. The torsion for the $\kappa$-Poincaré theory can also be non-zero, but it is zero for the MS DSR. The curvature for the $\kappa$-Poincaré theory and the MS DSR are zero. We will find that the non-zero torsion and curvature of the momentum space implies a non-commutative spactime which is tangent to this momentum space. Also, we show that the torsion for every Abelian DSR theory is zero at the origin of the momentum space. At the end, we will discus dual spacetime transformations for the $\kappa$-Poincaré theory and  MS-DSR. 
\end{abstract}

\maketitle

\tableofcontents

\section{Introduction}\label{intro}

Doubly special relativity (DSR) theories \cite{ame1, mag1, Wang:2013bta, Jafari:2011sg,Jafari:2020ywd}, have non-trivial geometries in the
 momentum space. In the context of relative locality theory \cite{ame3,ame4}, their momentum spaces are curved. This curvature
 is a consequence of the non-linearity in the additions of momenta. 

For a better understanding of these curved momentum spaces, we will compute the connection, torsion and the curvature for the $\kappa$-Poincaré theory \cite{Maj}, and the Magueijo-Smolin (MS) DSR \cite{mag1}. The  $\kappa$-Poincaré theory, and the MS DSR  are two main examples of the DSR theories. In fact, they are prototype examples of the DSRs in this two decade history of investigation into the character of DSR theories. Finding the connection, torsion and the curvature for the $\kappa$-Poincaré theory, and the MS DSR are important steps forward in this program.

Relative locality theory is based on some physical and some philosophical assumptions. One fundamental assumption is that we as observers live in phase space and spacetime is the tangent space to this phase space. Physical events occur in this phase space and there is no invariant
global spacetime. Every observer has their own spacetime and different observers construct different spacetimes, which are observer-dependent projections of phase space \cite{ame3}. The assumptions underlying DSR theories which have been obtained as deformations of special relativity include adding two additional principles: that the Planck energy is observer independent and at energies much smaller than the Planck energy conventional special relativity is true \cite{mag1}. Thus, finding fundamental relative locality parameters such as the connection, torsion and the curvature for the $\kappa$-Poincaré theory and the MS DSR are worthwhile. 

Amelino-Camelia \cite{ame5}, and Camonra's group \cite{cam1, cam2}  have investigated examples of the DSR theories and the $\kappa$-Poincar\'e theory in the context of relative locality theory. They have considered these matters at the first leading order of the Planck length. Here, we compute the connection, torsion and the curvature in all orders of the Planck length.

Expressions for the connection, torsion and the curvature in all orders of the Planck length, gives us a more complete picture of these quantities for the $\kappa$-Poincar\'e theory  and the MS DSR. Sometimes quantum gravity effects occur in second order or higher \cite{cam2}. Thus, this study complements Amelino-Camelia \cite{ame5}, and Camonra's group \cite{cam1, cam2} first order discussions.

An important paper by G. Gubitosi and F. Mercati has discussed the Hopf algebraic structures of the $\kappa$-Poincaré theory \cite{Gubitosi:2013rna}. Also, a paper by G. Gubitosi and S. Heefer \cite{Gubitosi:2019ymi} was devoted to the compatibility of the interacting $\kappa$-Poincaré theory with relative locality theory. These two papers are important studies in this direction. The Snyder model in the framework of the  relative locality theory has also been studied in \cite{Banburski:2013jfa, Mignemi:2016ilu, Ivetic:2017gte}.

The paper is organized in the following: in Sec~\ref{sec1} we give a brief history for assigning curvature to the momemtum space. In fact, we give some historical discussions for why and how the curved momentum space should be. In Sec.~\ref{Sec2}, we provide a brief introduction to relative locality theory. Then we compute the connection, torsion and curvature for the $\kappa$-Poincar\'e theory and MS DSR in the context of the relative locality in Sec.~\ref{Sec3} and Sec.~\ref{Sec4}, respectively. The physical implications of the connection, torsion and curvature in the momentum space are discussed in Sec. \ref{Sec5}, and finally we conclude in Sec. \ref{Conclusion}. Throughout we use natural units, $\hbar=c=1$.

\section{Curvature in the momentum space: why and how?} \label{sec1}

The first suggestion for a curved momentum space was given at the third decade of the past century by Max Born \cite{Born1938ASF}. 
By using the \emph{principal of reciprocity} which states that the laws of nature are invariant under  
\be    x_\mu   \rightarrow p_\mu ~~,~~~~  p_\mu   \rightarrow -x_\mu,\ee
he argued that the momentum space should be a curved space similar to the spacetime in Einstein's general relativity. 
Born proposed an equation similar to Einstein's general relativity 
\be       R_{\mu \nu}  - \frac{1}{2} R g_{\mu \nu} = - 8 \pi G T_{\mu \nu}, \ee
in the momentum space
\be       \Tilde{P}_{\mu \nu}  - \frac{1}{2}  \Tilde{P} \gamma_{\mu \nu} = - \Tilde{k} \Tilde{T}_{\mu \nu},\ee
where $\gamma_{\mu \nu}$ is the metric of the momentum space, $\Tilde{P}_{\mu \nu}$ is the Ricci curvature tensor in the momentum space and $\Tilde{P}$ is the corresponding scalar curvature. Also, $ \Tilde{T}_{\mu \nu}$  is the spacetime correspondence of the energy-momentum tensor $T_{\mu \nu}$, and $\Tilde{k} $ is a constant. But, Born's suggestion has not lead to good results for the quantization of gravity as discussed by Amelino-Camelia \cite{Amelino-Camelia:2012vpb}. 

The next attention to the curved momentum space was due to the 1960-1990 period and especially from Russian physicists  \cite{Golfand:1962kjf, Tamm:1965, Batalin:1989xu}. They were looking for a divergence free quantum field theory. 

Around the beginning of this century attention has been given to curved momentum space. Some of mathematical and 
conceptual papers by Shahn Majid are devoted to discussing the necessities for introducing curvature in the momentum
space \cite{Majid:1988we, Majid:1999tc, Majid:1999td}. He has argued that the curvature in position space implies non-commutativity in the momentum space. As an example when the position space is a 3-sphere which has momentum algebra $su_2$, the observables of the enveloping algebra $\Tilde{U}(su_2)$ will be satisfied in the relations
\be [p_i, p_j ] = \frac{1}{R}\epsilon_{ijk} p_k, \ee
in which $R$ is related to the radius of the curvature of the 3-sphere. By using Born reciprocity, we then have the possibility for curvature in momentum space. As another example, if the momentum space is a sphere with $m$ proportional to the radius of curvature then the noncommutation relations 
in position space will be 
\be [x_i, x_j ] = \frac{1}{m}\epsilon_{ijk} x_k. \ee
In Majid's approach the curved momentum space is alongside the noncomutativity in the spacetime \cite{Majid:1999tc}. 

There are some attempts which assumes 8-dimensional line elements of the phase space by combining the spacetime and the momentum space line elements 
\be ds^2 = dt^2 - dx^2 - dy^2 - dz^2 + \frac{1}{b^2}( dE^2 - dp_x^2 - dp_y^2 - dp_z^2 ), \ee
in which $b$ is a constant that depends on the Planck scale \cite{ Slawianowski, Low1993U31TW, Castro:2008zzc, Pavsic:2009xv, CastroPerelman:2020ysg}. Here the use of the Born reciprocity is for treating a 8-dimensional phase space and the goals of these approaches are mainly for extensions of general relativity.

In a similar way, we can generalize the Hamiltonian equations in the 8-dimensional curved phase space \cite{Barcaroli:2015xda, Relancio:2020rys}. This 
approach belongs to the Hamiltonian geometry category which is also related to Finsler geometry \cite{Pfeifer:2019wus}. Finsler geometry is a generalization of Riemann geometry with a position and velocity dependent metric. 
Starting from the free particle dispersion relation 
\be E^2= \textbf{p}^2 + m^2, \ee
and defining the Hamiltonian as 
\be    H(x, p) = m^2,      \ee
we can introduce any deformations from the usual quadratic form, e.g.  
 \be H(x, p) = p_0^2 - \textbf{p}^2 + l_p Q^{abc} p_a p_b p_c, \ee
in which $l_p$ is the Planck length and $ Q^{abc}$ is a matrix of numerical coefficients. Then we can define phase space metric as
\be  g^H_{\mu \nu} (x,p)= \frac{1}{2} \frac{\partial}{\partial p_\mu} \frac{\partial}{\partial p_\nu} H(x, p),  \ee
and by using a non-linear connection we can find the curvature of the momentum space \cite{Barcaroli:2015xda}.

In the first years of the twenty century it was shown that the DSR and the $\kappa$-Poincar\'e theories can be understood in terms of the curved momentum space as a subspace of the de Sitter space \cite{Kowalski-Glikman:2003qjp}. For continuation we consider the momentum space of a $\kappa$-Poincar\'e particle which is a four dimensional group manifold of a Lie group $AN(3)$ 
\cite{Kowalski-Glikman:2013rxa}. The generators  of the Lie algebra for this group satisfy 
 \be [X^0,X^i]= \frac{i}{\kappa} X^i. \ee 
  Matrix representations of this Lie algebra are 5-dimensional ones which one of them is the Abelian generator $X^0$ and the others are three nilpotent generators $X^i$. An group element of $AN(3)$  is 
  \be  u(p) = e^{ip_i X^i} e^{ip_0 X^0},\ee
  which acts on the 5-dimensional Minkowski space as a linear transformation. If this group element acts on 
  a special point $ (0, . . . , 0, \kappa) $, it will give a point in the 5-dimensional Minkowski space with coordinates  $(P_0, P_i, P_4)$ with 
    \be    P_0 = \kappa \sinh{\frac{p_0}{\kappa}}  + \frac{\textbf{p}^2}{2\kappa} e^{p_0/\kappa}, ~~~ P_i = p_i e^{p_0/\kappa},
~~~    P_4 = \kappa \cosh{\frac{p_0}{\kappa}}  - \frac{\textbf{p}^2}{2\kappa} e^{p_0/\kappa},\ee
 here $ p_\mu$ are the flat coordinates. It can be shown that 
  \be - P_0^2 + \textbf{P}^2 + P_4^2 = \kappa^2. \ee 
 Thus, the manifold of the group  $AN(3)$ is isomorphic to 4-dimensional de Sitter space. In the flat coordinates the line element  will be
\be  \label{KPionMetric} ds^2 = -dp_0^2 + e^{2p_0/\kappa} d\textbf{p}^2,\ee
which shows a curved momentum space.
For this metric we can calculate the distance function
\be C(p) = \kappa^2 \cosh^{-1}{\frac{P_4}{\kappa}}. \ee
Thus, the mass-shell condition will be 
\be \cosh{\frac{p_0}{\kappa} - \frac{\textbf{p}^2}{2\kappa} e^{p_0/\kappa}}=\cosh{\frac{m}{\kappa}}.
\ee

In a close similarity, we have $\kappa$-Minkowski deformations which are obtained from the deformed coordinates 
 \be     [x_i, x_j]=0, ~~~ [x_0, x_i]= \frac{i}{\kappa} x_i. \ee
The geometrical structure of the $\kappa$-Minkowski momentum space is interesting \cite{Lizzi:2020tci, Juric:2012xt, Juric:2013mma}. For example, combinations of the plane waves in these non-commutative coordinate will be given in a nontrivial way
 \be \exp{(ik_\mu x_\mu)}\exp{(iq_\mu x_\mu)}=  \exp{ \Big\{ i\frac{(k_0 + q_0)/\kappa}{e^{(k_0+ q_0)/\kappa} - 1}  
  \Big[ \Big(\frac{e^{k_0/\kappa} - 1}{k_0/\kappa} \Big) k_i + e^{-k_0/\kappa} \Big(\frac{e^{k_0/\kappa} - 1}
  {q_0/\kappa} \Big)q_i  \Big]x^i  + i(k_0 + q_0)x^0 \Big\} }.\ee
 This type of non-triviality shows that $\kappa$-Minkowski spacetime is associated with a curved momentum space. The metric for 
 this curved momentum space is like Eq.~(\ref{KPionMetric}), however we can find many different metrics by using other methods such as embedding in 5-dimensional metric preserving groups, Riemannian hyperbolic momentum space, and two-time hyperbolic space \cite{Lizzi:2020tci}. 

As seen here the constructions of the curved momentum space are not unique, since several inequivalent momentum space geometries can be introduced. The curvature that we are discussing here has originated from the non-linearity in the combinations of the momentums as given in relative locality theory.

\section{Relative Locality theory} \label{Sec2}

For orientation, we briefly introduce the relative locality theories which are based on the nonlinear combinations of momenta. In these theories, momentum 
space $ \textsf{P}$ is assumed to have a nontrivial geometry, and we give some definitions for the connection, curvature and the torsion in the momentum space manifold \cite{ame3, ame4}. The addition rule is defined by a $C^{\infty} $ map
\be  \label{Addition}  \oplus: \textsf{P} \times \textsf{P} \longrightarrow \textsf{P}, ~~~~
 (p,q)\rightarrow p \oplus q, \ee
 such that it has an identity element $0$: 
 \be    p \oplus 0= 0 \oplus p=p,\ee
 and an inverse element $\ominus$:
  \be  \ominus: \textsf{P} \times \textsf{P} \longrightarrow \textsf{P}, ~~~~  
    \ominus p \oplus p= p \ominus p=0. \ee

In special relativity the momentum space is the Minkowski linear momentum space and the combinations of 4-momenta is the usual vector addition $p'=p+q$.  We also have the usual inverse $p-p=0$. 

For the combination rule Eq.~(\ref{Addition}) one could define a \emph{connection} by
\be \Gamma_{c}^{ab}(0)=-\frac{\partial}{\partial p_a} \frac{\partial}{\partial q_b}
(p \oplus q)_c|_{q,p=0}.\label{connection}\ee
\emph{Torsion} of this combination is defined by
\be T_c^{ab}(0)=-\frac{\partial}{\partial p_a}
\frac{\partial}{\partial q_b}[(p\oplus q)_c-(q\oplus p)_c]|_{q,p=0}.\label{torsion}\ee
Using this connection Eq.~(\ref{connection}), one can define a \emph{curvature} for the 4-momentum space
\be \label{curvture} R^{abc}_{~~~d}(0) = 2\frac{\partial}{\partial p_{[a}} \frac{\partial}{\partial q_{b]}}
\frac{\partial}{\partial k_c}\Big((p\oplus q)\oplus k-p\oplus(q\oplus
k)\Big)_d\Big|_{p,q,k=0}, \ee
where as usual, the bracket denotes anti-symmetrization. Curvature could be interpreted as a lack of associativity of the combination
rule.   

By considering translation in momentum space we can find the connection, torsion and the curvature in an arbitrary point in momentum space.
The translated composition rules of the momenta at the arbitrary point $k$ in the momentum space are
 \be \label{TranslSum} p\oplus_k q=  k\oplus((\ominus k\oplus p)\oplus(\ominus k\oplus q)),  \ee 
where the identity for this product is at $0_k = k$. By rewriting Eq.~(\ref{connection}), Eq.~(\ref{torsion}) and Eq.~(\ref{curvture}) by the translated composition rule $\oplus_k$, and using them, we can compute the connection, torsion and the curvature tensors away from zero momenta. 

\begin{figure}[ht]
\centering 
\includegraphics[width=2.5in]{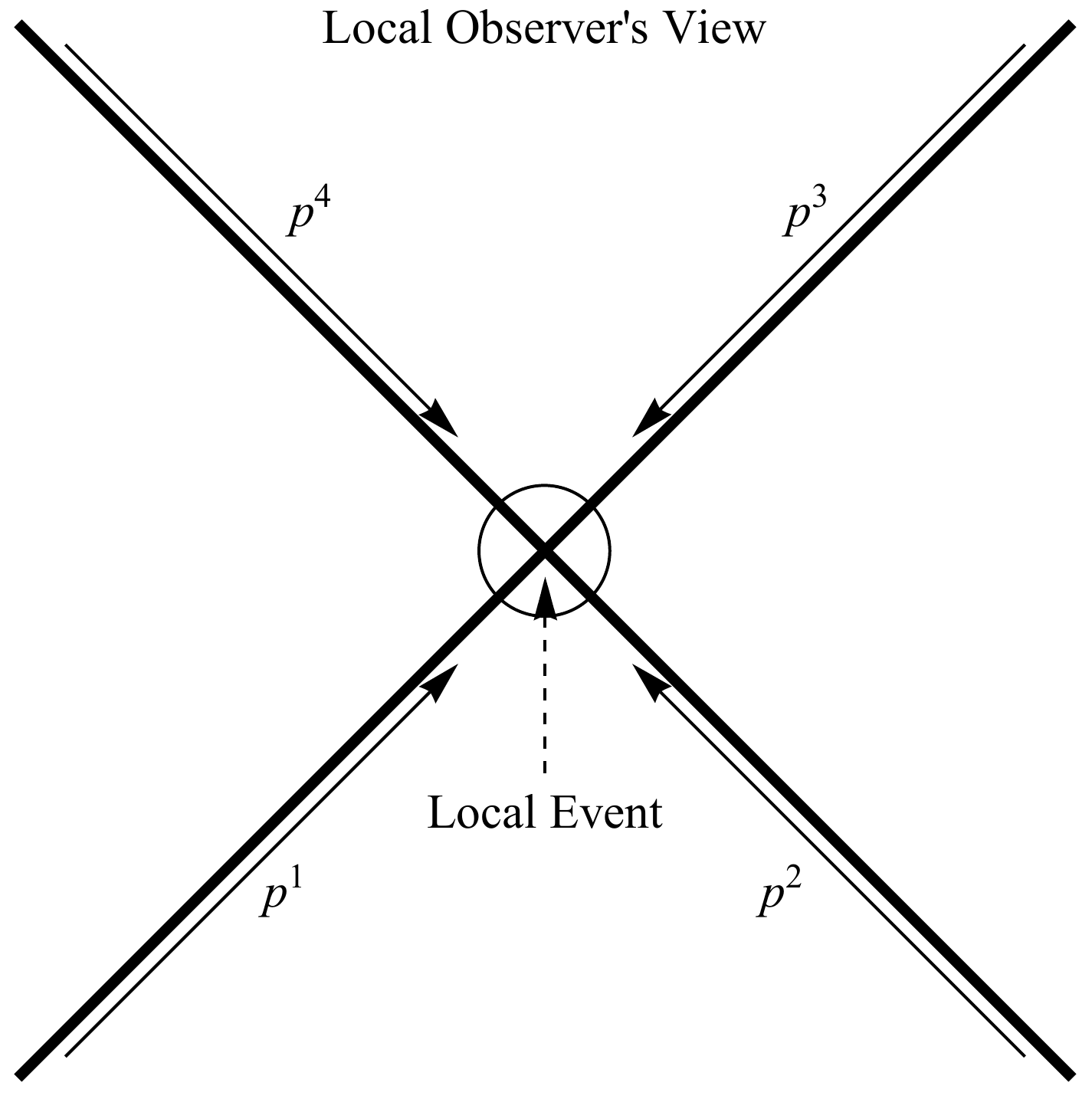}\;\;
\includegraphics[width=2.65in]{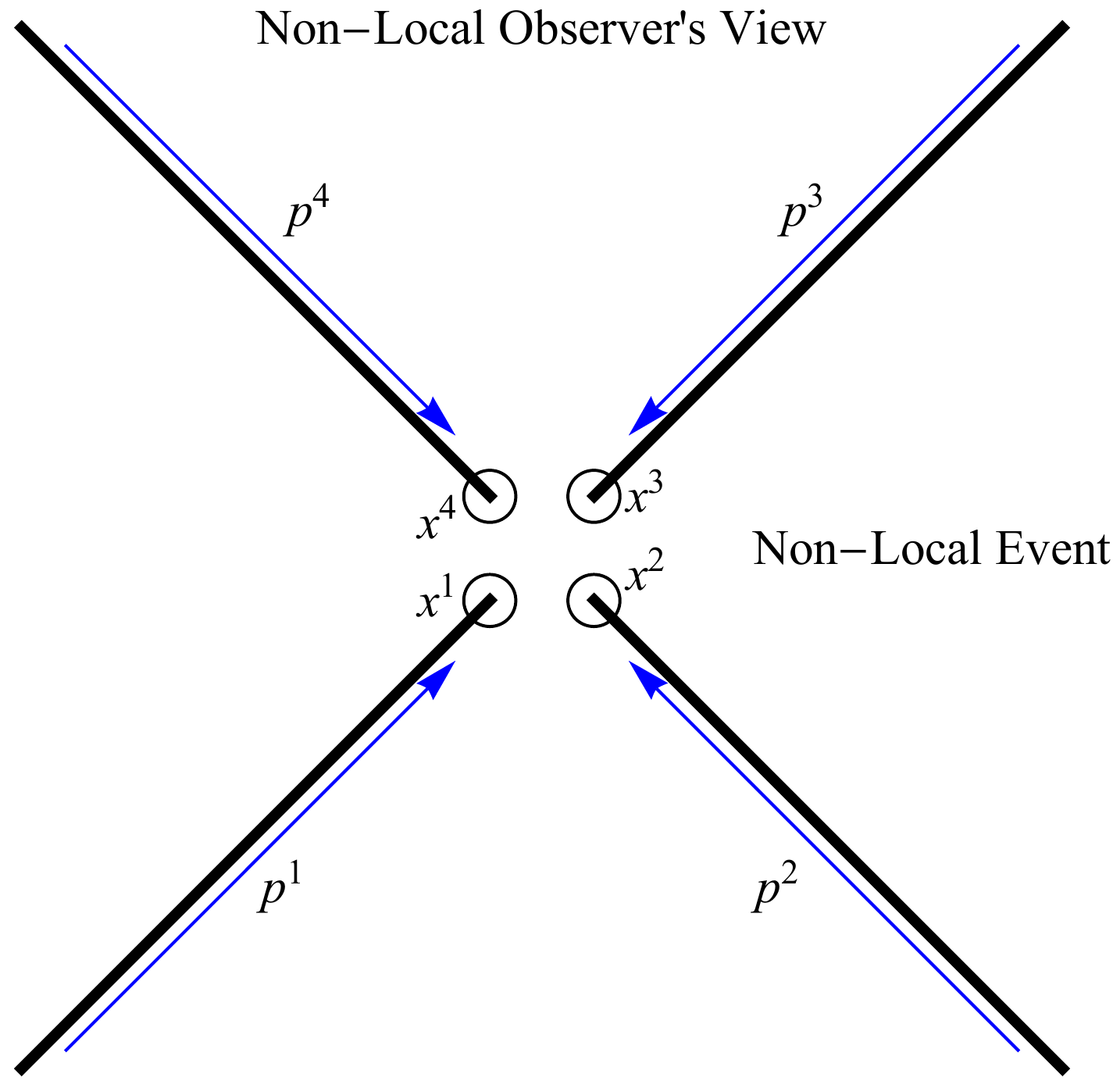}
\caption{ The energy dependency of the spacetime for the distant observer in relative locality implies non-locality for the event in the right figure. For the distant observer, the projections of the ends of the worldlines no longer meet. But, for the event at the origin of the observer’s coordinate system, the event is like usual special relativity and has the usual local description. See also Figure 2 from \cite{ame4}. } 
\label{fig1} 
\end{figure}

In the relative locality theory, the mass of a particle is interpreted as the geodesic distance from
the origin of the momentum space
\be C(p)=m^2 ,\ee 
which gives the dispersion relation, and we have $C(p)= \hat{g}_{\mu \nu} p^\mu p^\nu$. In other words, one ascribes a metric $\hat{g}_{\mu \nu}$ to
the momentum space, which is not in general the same as the Minkowski metric. In this theory, the metric of the spacetime $g_{\mu \nu}$ can also depend on the  energy and momentum. Different observers see different spacetimes which means coincidences of events are not the same for all observers. This matter has been illustrated in Fig.~(\ref{fig1}).

%%%%%%%%%%%%%%%%%%%%%%%%%%%%%
%%%%%%%%%%%%%%%%%%%%%%%%%%%%%  
%%%%%%%%%%%%%%%%%%%%%%%%%%%%%
%%%%%%%%%%%%%%%%%%%%%%%%%%%%%
%%%%%%%%%%%%%%%%%%%%%%%%%%%%%
%%%%%%%%%%%%%%%%%%%%%%%%%%%%%

\section{Connection, Torsion and Curvature for the $\kappa$-Poincar\'e theory} \label{Sec3}

\subsection{At the origin of the momentum space}
We consider the $\kappa$-Poincar\'e compositions of the momenta with Majid-Ruegg connection \cite{Maj}. For the compositions of momenta at the origin of the momentum space we have 
\be  \label{CompKapa}  \left\{\begin{array}{cl}   (p\oplus q)_0 &=  p_0 + q_0, \\\\
(p\oplus q)_i &=  p_i + e^{-lp_0} q_i, \end{array}\right.\ee 
from which we can compute the connection, torsion and the curvature at the origin of the momentum space. The connections are 
\be\left\{\begin{array}{cl}   \Gamma^{ab}_0(0)= 0, \\\\
 \Gamma^{ab}_i(0)= l \delta^a_0 \delta^b_i, \end{array}\right.\ee 
 and the torsions are 
 \be\left\{\begin{array}{cl}   T^{ab}_0(0)= 0, \\\\
 T^{ab}_i(0)= l \delta^{[a}_0 \delta^{b]}_i. \end{array}\right.\ee
 
 For finding curvature we compute combinations of three momenta and associativity
of the combination rules
\be \Big[ (p\oplus q)\oplus k \Big]_i= p_i + e^{-lp_0} + e^{-l(p_0 + q_0)} k_i, \ee
which is equal with $ \Big[  p\oplus(q\oplus k) \Big]_i $. Thus, using Eq.~(\ref{curvture}), the curvature of the $\kappa$-Poincar\'e at the 
origin will be zero 
\be      R^{abc}_{~~~d}(0)=0.        \ee

 \subsection{At the arbitrary point $k^{\mu} $ of the momentum space }
 
Antipodes for composition rules Eq.~(\ref{CompKapa}) are 
\be\left\{\begin{array}{cl}   (\ominus p)_0 &= -p_0, \\\\
 (\ominus p)_i &= - e^{-lp_0}p_i, \end{array}\right.\ee 
 by using them and Eq.~(\ref{TranslSum}), we can find the translated composition rules 
\be \label{TrComRul}  \left\{\begin{array}{cl}   (p \oplus_k q)_0 &=p_0 + q_0, \\\\
 (p \oplus_k q)_i &= p_i + e^{l(k_0 - p_0)} q_i + k_i (  1- e^{-2lk_0} - e^{-l(k_0+ p_0)}    ). \end{array}\right.\ee 
 Thus, the translated connections
\be \label{ }  \Gamma_{c}^{ab}(k)=-\frac{\partial}{\partial p_a} \frac{\partial}{\partial q_b}
(p \oplus q)_c|_{q,p=k},  \ee
 can be obtained from the translated combination rules as
\be\left\{\begin{array}{cl}   \Gamma^{ab}_0 (k)=0, \\\\
 \Gamma^{jk}_i (k) = l \delta^j_0 \delta^k_i. \end{array}\right.\ee

The  translated torsions are defined by
\be \label{torsionWITHk} T_c^{ab}(k)=-\frac{\partial}{\partial p_a}
\frac{\partial}{\partial q_b}[(p\oplus q)_c-(q\oplus_k p)_c]|_{q,p=k},\ee
and by using Eq.~(\ref{TrComRul}), they will be given by 
\be  \label{torsionKP} \left\{\begin{array}{cl}   T^{ab}_0 (k) =0, \\\\
 T^{jk}_i (k)= l \delta^{[j}_0 \delta^{k]}_i. \end{array}\right.\ee
 
 By using the translated combination rules Eq.~(\ref{TrComRul}), we obtain  the composition of three momenta
     \be \Big[p \oplus_k ( q\oplus_k k) \Big]_0 =p_0 + q_0- k_0 = \Big[(p \oplus_k q )\oplus_k k \Big ]_0, \ee
  \be  \Big[p \oplus_k ( q\oplus_k k) \Big]_i = p_i + e^{l(k_0 - p_0)} q_i + k_i \Big(  1- e^{-2lk_0} - 2 e^{-l(k_0+ p_0)} 
  +   e^{l(k_0 - p_0)} - e^{-l(p_0- q_0)}  +  e^{ l(2k_0 -p_0-q_0) }  \Big) , \ee
   and 
  \be \Big[(p \oplus_k q) \oplus_k k ) \Big]_i = p_i + e^{l(k_0 - p_0)} q_i + k_i \Big(  2- 2e^{-2lk_0} - e^{-l(k_0+ p_0)} 
+ (e^{l(k_0-1)}-1) e^{ -l( p_0+ q_0- k_0) }  \Big) ,\ee 
and the non-associativity is 
\be         \Big[(p \oplus_k q )\oplus_k k \Big ]_i -  \Big[p \oplus_k (q \oplus_k ) \Big]_i =  k_i \Big(  1- e^{-2lk_0} +  e^{-l(k_0+ p_0)} +
e^{ -l(p_0+q_0) }   +  (e^{l(k_0-1)}-1) e^{ -l( p_0+ q_0- k_0)}-  e^{ l(2k_0 -p_0-q_0) } \Big).       \ee
After a triple derivative we find
\bea  
\frac{\partial}{ \partial p_a}   \frac{\partial}{ \partial q_b}   \frac{\partial}{ \partial k_c}
\Big[(p \oplus_k q )\oplus_k k  -  p \oplus_k (q \oplus_k ) \Big]_i = &-& l^3 \Big(2e^{l(k_0)}-1 \Big) e^{ -l( p_0+ q_0- k_0)} \delta^a_0 \delta^b_0 \delta^c_0 k_i\\ \nonumber
&-&l^3 \Big(e^{l(k_0)}-1 \Big) e^{ -l( p_0+ q_0- k_0)} \delta^a_0 \delta^b_0 \delta^c_i  \\ \nonumber
&-&l^3  e^{ -l( p_0+ q_0)} \delta^a_0 \delta^b_0 \delta^c_0 k_i   \\ \nonumber
&-& 2 l^3 e^{ -l(2k_0- p_0- q_0)} \delta^a_0 \delta^b_0 \delta^c_0 k_i \\ \nonumber
&-&l^2  e^{ -l(2k_0- p_0- q_0)} \delta^a_0 \delta^b_0 \delta^c_0 k_i. \nonumber
\eea
This term is symmetric with respect to $p$ and $q$. Thus, the curvature which are given by 
 \be  \label{CurAtk}      R^{abc}_{~~~d} (k)= \frac{\partial}{ \partial p_{[a}}   \frac{\partial}{ \partial q_{b]}}   \frac{\partial}{ \partial k_c}
  \Big[(p \oplus_k q )\oplus_k k  -  p \oplus_k (q \oplus_k ) \Big]_i\Big|_{q,p=k}, \ee 
  will be zero at every point of the momentum space for the $\kappa$-Poincar\'e theory 
   \be \label{CurvKP} R^{abc}_{~~~d} (k)=0.\ee

\section{Connection, Torsion and Curvature for the MS DSR} \label{Sec4}

Abelian DSR theories have commutative compositions of the momenta and can be mapped to special relativity. The Amelino-Camelia \cite{ame1}, and the MS \cite{mag1} DSR theories are Abelian. For any Abelian DSR theory, we have a relation between $p_\mu$ and $\pi_\mu$ as
    \be  \pi_\mu= U(p_\mu)p_\mu.  \ee
 The $ U(p_\mu)$ is an operator which map the Abelian DSR theory to special relativity, and the $\pi_\mu$ is the special relativistic variable. 
 By using the composition of the momenta for special relativistic variables $\pi_\mu$, we can find the composition of the momenta for $p_\mu$ variables as 
      \be    \label{momenAbelComp} (p\oplus q)_{\mu}= U^{-1} \Big(  U(p_\mu)p_\mu + U(q_\mu)q_\mu \Big) \Big(  U(p_\mu)p_\mu + U(q_\mu)q_\mu \Big) .                 \ee
    
    This composition rule is symmetric with respect to $p$ and $q$, and from Eq.~(\ref{torsion}) we find that the torsion for the Abelian DSR theories is zero at the origin of the momentum space. This is an interesting result, as we know most of the well known DSR theories are Abelian. Thus, we expect non-zero torsion only in the non-Abelian DSR theories. 
    
\subsection{At the origin of the momentum space}

The composition rules of the momentums for MS DSR at the origin of the momentum space are 
     \be  \label{comp_gen_MS}  (p\oplus q)_{\mu} = \frac{ (  1- lq_0  ) p_{\mu} +  (  1- l p_0 ) q_\mu }{1- l^2 p_0 q_0}.   \ee 
     Also by using these rules and from Eq.~(\ref{connection}), we find that the connection at the origin of the momentum space is
     \be    \Gamma^{ab}_c (0)= l (\delta^a_c \delta^b_0 +  \delta^a_0 \delta^b_c  ).        \ee 
     This expression is symmetric with respect to $p$ and $q$. Thus, the torsion tensor in the origin of the momentum space which is given by Eq.~(\ref{torsion}) will be zero 
             \be   T_c^{ab}(0)=0.  \ee

     The composition rules for the three momenta by using of Eq.~(\ref{comp_gen_MS}) are given by 
     \be      \Big[ (p\oplus q)\oplus k \Big]_\mu=  \frac{ (1- lq_0)(1-lk_0)p_\mu +(1- lp_0)(1-lk_0)q_\mu + (1- lp_0)(1-lq_0)k_\mu }
    {1- l^2 p_0q_0  - l^2 p_0 k_0 - l^2 q_0 k_0 + 2l^3 p_0q_0 k_0    },\ee  which are the same with  $ \Big[  p\oplus(q\oplus k) \Big]_\mu$. Thus, 
    by using  Eq.~(\ref{curvture}), curvature for the MS DSR at the origin is zero,
     \be     R^{abc}_{~~~d}(0)=0.       \ee

          \subsection{At the arbitrary point $k^{\mu} $ of the momentum space }
          
          The antipodes  for the MS DSR are
             \be    (\ominus p)_\mu= - \frac{p_\mu }{1- 2 l p_0},  \ee 
             and by using Eq.~(\ref{TranslSum}),  the compositions rules will be 
               \be (p \oplus_k q)_\mu= \frac{ (1-lq_0)(1-lk_0 ) p_\mu + (1-lp_0)(1-lk_0 ) q_\mu - (1-lp_0)(1-lq_0 ) k_\mu }
            {1-  2 l k_0 + l^2 k_0 p_0 + l^2 k_0 q_0 -l^2 p_0 q_0 }.\ee 
            Using these rules, we find the connection at point $k^\mu$ as
            \be   \Gamma^{ab}_c(k)= \frac{l ( \delta^a_c \delta^b_0 +  \delta^a_0 \delta^b_c  )}{1 - lk_0},  \ee 
            and also the torsion  will be zero 
            \be \label{torsionMS}  T_c^{ab}(k)=0.  \ee 
            
            Here, we have an interesting result that 
                 \be     p \oplus_k k= p.     \label{poplusk} \ee
           By using this property, we find that the compositions of three momenta will be
            \be   ( p \oplus_k q) \oplus_k k = p \oplus_k q=   p \oplus_k (q \oplus_k k) .\ee
            Thus, the curvature for the MS DSR at any arbitrary point of the momentum space which is given by Eq.~(\ref{CurAtk}) will be zero
            \be \label{CurvMS} R^{abc}_{~~~d}(k)=0.\ee

\section{ Physical implications of the connection, torsion and curvature} \label{Sec5}

As given before in the main papers of relative locality \cite{ame3, ame4}, the results for the connection, torsion and the curvature have physical implications. The connections enter into the conservation laws of the momenta
 \be  \label{Conservrule}  P_\mu^{total}= \sum_I p_\mu^I -  l_p \sum_{I < J}   \Gamma_\mu ^{\nu \alpha} p_\nu^I  p_\alpha^J.  \ee  
  Here $ J $ is the number of the  particles which interact with the $I$ particles. This formula is a new expression for the Judes-Visser 
 conservation rules of the momenta in the DSR theories in terms of the relative locality parameters \cite{Judes:2002bw}.  

 The connections can also specify the amount of non-locality for the observers of an interaction. If the first observer of an interaction has been separated from the second observer by a vector $b^\mu $, then the worldlines of the particles will
 be translated by the amounts
  \be     \label{NonLocalityAmount}   \Delta x_I^\mu =  b^\mu + l_p  b^\mu \sum_{I > J}  \Gamma_\mu ^{\nu \alpha}  p_\alpha^J.\ee 
  This is the relative locality effect which we explained before in Section \ref{Sec2}, and it has been illustrated in Fig.~(\ref{fig1}). 
  
  Howver, the curvature for the $\kappa$-Poincar\'e theory and for MS-DSR are zero, but conections are not zero in general. Thus, the conservation laws Eq.~(\ref{Conservrule}) remian nonlinear, also non-localities  Eq.~(\ref{NonLocalityAmount}) happen for distant observers in the interactions.

  \section{Tangent spacetime}
For every momentum space  $ \textsf{P}$, we have a tangent spacetime. Poisson bracket for the coordinate variables $x^a$ of this tangent spacetime \cite{ame3}, is given by 
  \be \label{Poisson}  \{x^a, x^b \}= T^{ab}_d x^d + R^{abc}_d p_c x^d.    \ee
  
  \subsection{ Tangent spacetime for $\kappa$-Poincar\'e theory }
  
   If we put the values of torsions  Eq.~(\ref{torsionKP})  and  curvatures Eq.~(\ref{CurvKP}) for the $\kappa$-Poincar\'e theory in the Eq.~(\ref{Poisson}) we find the Poisson brackets for the tangent spacetime variables of the $\kappa$-Poincar\'e theory as
    \be      \{x^a, x^b \}= l \delta^{[a}_0 \delta^{b]}_d x^d.   \ee
    Thus, we have 
      \be  \label{PosBraketKpioX} \{x^0, x^i \}= l x^i, \ee 
      which shows that the tangent spacetime  of the $\kappa$-Poincar\'e theory is not a commutative spacetime.
      
      In fact, Eq.~(\ref{PosBraketKpioX}) is a  part of the Poisson brackets for phase space in  $\kappa$-Poincar\'e  algebra.
      There are some approaches for finding dual theories for the $\kappa$-Poincar\'e theory and also the DSR \cite{Magpantay:2010zz,Magpantay:2010np}. Here, dual means spactime counterpart. We guess transformations for daul
      spactime shoud be similar to 
      \be  x'^a= H^a_b (\Lambda) x^b  + lx^a ,    \ee
      in which $H^a_b(\Lambda) $ is a function of $\Lambda$, and $ \Lambda $ is the Lorentz transformations matrix. Finding  $H^a_b$ can be a continuation of this paper.
      
  \subsection{ Tangent spacetime for MS-DSR }
  
   For the MS-DSR we use the values of the torsions  Eq.~(\ref{torsionMS})  and  curvatures Eq.~(\ref{CurvMS}) in the Eq.~(\ref{Poisson}) and we found that 
          \be \label{MSDSRPoisson}  \{x^a, x^b \}= 0, \ee 
          which shows that the tangent spacetime  of the MS-DSR is a commutative spacetime. 
          
\subsection{ Transformations for the Tangent space time of MS-DSR }
  For this  commutative dual  spacetime we can find transformations which relates  $x^a$ to the transformed $x'^a$ coordinates. These transformations has been obtained before by J. Maguijo and his colleagues \cite{Kimberly:2003hp}. 
  The MS-DSR is a  nonlinear representation of the Lorentz group
     \be K^i = U^{-1} L^i_0 U, \ee
     in which  $U(E,P)$ acts on $ p_a $ and we have  
    \be   U_a(E, p_i) = (Ef(E), p_i g(E)), \ee
    here, $f(E) $ and $ g(E) $ are arbitrary functions of energy. In fact, these functions  are obtained from a given modified dispersion relation. 
    
    For obtaining transformations for dual spacetime we should put some conditions. Here, we assume that $ U^a(t, x^i)$ acts on $x^a$  in a covariant way  
    \be U_a(p)U^a(x) = p_ax^a, \ee
    then $ U^a(t, x^i)$  will be
        \be   U^a(t, x^i) = ( \frac{t}{f(E)}, \frac{x^i}{ g(E)}). \ee
    Thus, the dual space time transformations will be

 \be t' = \gamma \Big(t - vx \frac{f(E)}{g(E)} \Big) \frac{f(E')}{f(E)}, \\~~~~
 x' = \gamma \Big(x - vt \frac{g(E)}{f(E)} \Big) \frac{g(E')}{g(E)}.\ee
 
 For the MS-DSR we have 
 \be  f(E)=g(E)= \frac{1}{1-lp_0}, \ee 
 and the dual space time transformations will be 
  \be t' = \gamma (t - vx  )  [1 + (\gamma - 1)l E - \gamma l v p ], \\~~~~
 x' = \gamma (x - vt )[1 + (\gamma - 1)l E - \gamma l v p ].\ee

\section{Conclusion}\label{Conclusion}

For the $\kappa$-Poincar\'e theory and the MS DSR, the connections can be non-zero. The torsion for the $\kappa$-Poincar\'e theory can also be non-zero, but because of the symmetric connection the torsion for the MS DSR is zero. The curvatures for the $\kappa$-Poincar\'e theory and the MS DSR are zero at the every point of the momentum space. Also, we showed that the torsion for the Abelian DSR theories is zero at the origin of the momentum space. We propose that curvature for every Abelian DSR theory should be zero at least at the origin of the momentum space. 

Continuation of this work includes deriving the general formulas for the curvature of the Abelian DSR theories. Finding general formulas for the non-Abelian DSR theories are intractable.

This study has lead to a better understanding of the DSR theories in the context of relative locality theory, and  provides formulas for the connection, torsion and the curvature for the $\kappa$-Poincar\'e theory and the MS DSR easily accessible. In addition, we hope that these results help clarify the non-trivial geometry of the DSR theories.

\section{ Acknowledgment}
This research is funded by the Science Committee of the Ministry of Science and Higher Education of the Republic of Kazakhstan Program No. BR21881880.

\textbf{Data Availability Statement:} No Data associated in the manuscript.

\bibliography{main}

\end{document}